\documentclass[journal]{IEEEtran}
\usepackage{amsmath}
\usepackage{amssymb}
\usepackage{amsthm}
\usepackage{amsfonts}
\usepackage{epsfig}
\usepackage{psfrag}
\usepackage{graphicx}
\usepackage{mathrsfs}
\usepackage{color}
\usepackage{cite}
\usepackage[colorlinks=true,linkcolor=black,anchorcolor=black,citecolor=black,filecolor=black,menucolor=black,runcolor=black,urlcolor=black]{hyperref}
\usepackage{multirow}

\usepackage{physics} 
\usepackage{empheq}
\usepackage{authblk}
\usepackage[normalem]{ulem} 
\usepackage{mathtools} 

\usepackage{algorithm}
\usepackage{algpseudocode}

\ifCLASSOPTIONcompsoc
\usepackage[caption=false,font=normalsize,labelfon
t=sf,textfont=sf]{subfig}
\else
\usepackage[caption=false,font=footnotesize]{subfig}
\fi

\theoremstyle{definition}

\newtheorem{remark}{Remark}

\title{Vertical Federated Edge Learning with Distributed Integrated Sensing and Communication}
\author{Peixi Liu, Guangxu Zhu, \textit{Member, IEEE}, Wei Jiang, \textit{Member, IEEE},\\ Wu Luo, \textit{Member, IEEE}, Jie Xu, \textit{Member, IEEE}, and Shuguang Cui, \textit{Fellow, IEEE}
\thanks{Peixi Liu is with State Key Laboratory of Advanced Optical Communication Systems and Networks, School of Electronics, Peking University, China, and Shenzhen Research Institute of Big Data, Shenzhen, China (e-mail: liupeixi@pku.edu.cn). \textit{(Corresponding authors: Guangxu Zhu and Wei Jiang.)}} 
\thanks{Guangxu Zhu is with Shenzhen Research Institute of Big Data, Shenzhen, China (e-mail: gxzhu@sribd.cn).}
\thanks{Wei Jiang and Wu Luo are with State Key Laboratory of Advanced Optical Communication Systems and Networks, School of Electronics, Peking University, Beijing, China (e-mail: jiangwei, luow@pku.edu.cn).}
\thanks{Jie Xu is with the School of Science and Engineering (SSE), the Future Network of Intelligence Institute (FNii), and the Guangdong Provincial Key Laboratory of Future Networks of Intelligence, The Chinese University of Hong Kong (Shenzhen), Shenzhen, China (e-mail: xujie@cuhk.edu.cn)}
\thanks{Shuguang Cui is with the School of Science and Engineering (SSE), the Future Network of Intelligence Institute (FNii), and the Guangdong Provincial Key Laboratory of Future Networks of Intelligence, The Chinese University of Hong Kong (Shenzhen), and Shenzhen Research Institute of Big Data, Shenzhen, China. He is also with Peng Cheng Laboratory (e-mail: shuguangcui@cuhk.edu.cn).}
}

\begin{document}
%
%
	\maketitle

\begin{abstract}
	This letter studies a vertical federated edge learning (FEEL) system for collaborative objects/human motion recognition by exploiting the distributed integrated sensing and communication (ISAC). In this system, distributed edge devices first send wireless signals to sense targeted objects/human, and then exchange intermediate computed vectors (instead of raw sensing data) for collaborative recognition while preserving data privacy. 
	To boost the spectrum and hardware utilization efficiency for FEEL, we exploit ISAC for both target sensing and data exchange, by employing dedicated frequency-modulated continuous-wave (FMCW) signals at each edge device. Under this setup, we propose a vertical FEEL scheme for realizing the recognition based on the collected multi-view wireless sensing data. In this scheme, each edge device owns an individual local L-model to transform its sensing data into an intermediate vector with relatively low dimensions, which is then transmitted to a coordinating edge device for final output via a common downstream S-model. By considering a human motion recognition task, experimental results show that our vertical FEEL based approach achieves recognition accuracy up to 98\% with an improvement up to 8\% compared to the benchmarks, including on-device training and horizontal FEEL.
\end{abstract}


\section{Introduction}

Edge intelligence has emerged as a promising technique for beyond-fifth generation (B5G) and sixth generation (6G) wireless networks, which aims to exploit massive distributed data and computing power at wireless edge to enable abundant intelligent applications such as auto-driving, smart home, and object/activity recognition \cite{Park2019Edge}. 
Among others, federated edge learning (FEEL), a framework derived from federated learning (FL) at the network edge, is particularly appealing, in which artificial intelligence (AI) tasks are distributed over separate edge devices by leveraging their data locally with privacy preservation \cite{Zhu2020CM-edge}.
There are two widely used FEEL paradigms, namely horizontal and vertical FEEL, respectively.
While the horizontal FEEL (H-FEEL) focuses on the ``sample-partition'' scenario when different edge devices share the same feature space but have different sample spaces, the \textit{vertical FEEL} (V-FEEL) targets the ``feature-partition'' scenario when these edge devices have distinct feature spaces (with possibly partial overlapping) but share the same sample space \cite{Yang2019FL-overview}.
In the literature, there have been various prior works investigating H-FEEL from different perspectives such as communication and computation \cite{Lim2020Edge-survay}. 
However, how to design efficient FEEL (especially the V-FEEL) to achieve the highly-accurate sensing has not been well investigated yet.

\begin{figure}[!t]
	\centering
	\includegraphics[width=0.48\textwidth]{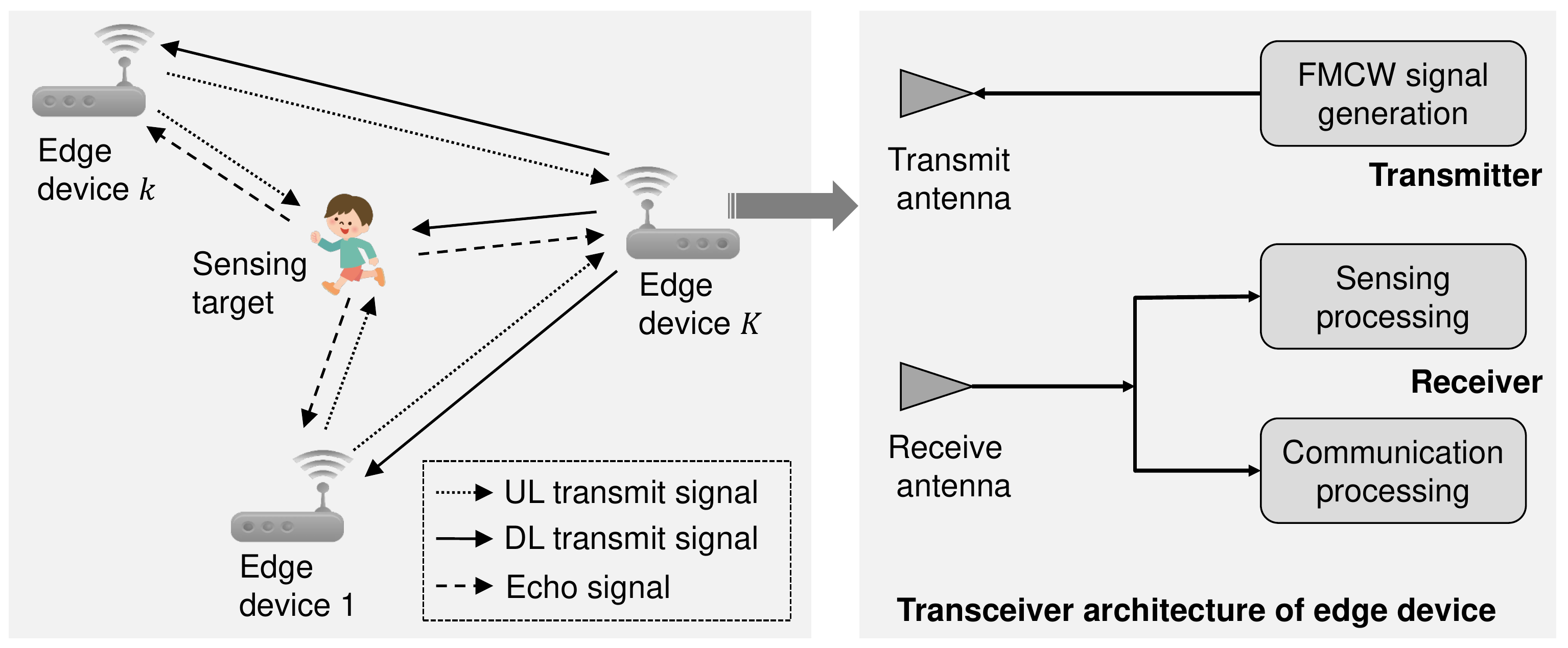}
	\caption{Vertical federated edge learning system with $K$ edge devices, where each device is equipped with a wireless sensor and edge device $K$ is selected as the coordinator.}
	\label{fig:system}
\end{figure}

On the other hand, wireless sensing has emerged as a new promising sensing technique as it can work under any weather or light conditions and its collected information is less sensitive than commonly used camera sensors \cite{Haque2020Nature}.
Wireless sensing can also be easily integrated with wireless communications to enable the so-called \textit{integrated sensing and communication} (ISAC), which can significantly enhance the spectrum and hardware utilization efficiency \cite{Liu2021arXiv-ISAC}.
By utilizing advanced machine learning approaches, such as support vector machine (SVM) and deep learning (DL), wireless sensing has been successfully applied in applications such as human motion recognition, sleep monitoring, and gait recognition \cite{Gurbuz2019SPM_SensingDL}.
In particular, it has been shown in \cite{Chen2018GRSL-multi-view} that jointly exploiting multiple distributed sensing nodes to acquire data from different aspect angles can significantly enhance the sensing performance.
This is due to the fact that sensing nodes placed at different locations can obtain unique observations related to the same target from different views, which can thus provide diversified features to describe the same target.
This fact has motivated the development of several existing \textit{distributed wireless sensing} (DWS) designs.
For instance, in \cite{Chen2018GRSL-multi-view}, the raw data of different sensing nodes are fused at a centralized node, which is then fed into the AI model.
However, this may lead to severe data privacy issue, especially when the sensing nodes belong to different entities with self interests. 
Recently, there was one prior work \cite{Yang2022FL-multi-radar} investigating the H-FEEL with DWS, but the H-FEEL mechanism targeting sample-partition scenario cannot fit well with the feature-partition nature of the DWS applications as it failed to utilize the feature diversity provided by the multi-view sensing observations, thus leading to degraded sensing accuracy.

To resolve the above issues, this letter proposes a new V-FEEL scheme for DWS-based recognition, in which distributed edge devices use wireless signals to sense targeted objects/human, and then exchange their computed intermediate vectors (instead of raw data in centralized learning \cite{Chen2018GRSL-multi-view} or local parameters of a common model in H-FEEL \cite{Yang2022FL-multi-radar}) for collaborative recognition while preserving data privacy. In this V-FEEL scheme, each edge device owns a local L-model to transform its input sample to an intermediate vector with relatively low dimensions, which is then transmitted to a coordinating edge device for the final output via a common downstream S-model. Furthermore, to enhance the communication and sensing efficiency of V-FEEL, we further exploit ISAC designs by employing frequency-modulated continuous-wave (FMCW) signals for the dual purpose of target sensing and data exchange. To the best of our knowledge, this work is the first to apply V-FEEL and ISAC into DWS. 
Furthermore, we demonstrate the performance of our proposed V-FEEL scheme via a concrete sensing task of human motion recognition and show its superiority over other benchmarks such as on-device training and H-FEEL \cite{Yang2022FL-multi-radar}.

\section{System Description}

We consider a FEEL-DWS system including $K$ edge devices in $\mathcal{K} \triangleq \{1,\ldots,K\}$ distributed at different orientations of the target, each of which has a wireless sensor that transmits and receives radio signals for both sensing and communicating with other devices, as shown in Fig. \ref{fig:system}. 
The edge devices collect the sensing data, which are then used in the training and inference process of the V-FEEL, by processing their received echo signals.
\color{black}
Since the dataset is generated at each edge device, the labels are held by the edge devices. To avoid edge devices from sharing label information to other parties, one edge device, e.g., device $K$, is selected as a coordinator\footnote{\textcolor{black}{In V-FEEL, the training loss needs to be computed at the coordinator. If any other party, e.g., an edge server, is selected as the coordinator, the labels need to be transmitted to the edge server in order to compute the training loss, which may lead to privacy issue.}}.
\color{black}
Besides sensing the target, the coordinator device $K$ is also responsible for communicating with the rest of edge devices for necessary information exchange required for training and inference in the V-FEEL.  
The details of the V-FEEL training and inference processes will be elaborated in Section \ref{sec:VFL}. 
In the following, we introduce the signal model and the signal preprocessing in the DWS system.


\subsection{Signal model for ISAC}

\begin{figure}[!t]
	\centering
	\includegraphics[width=0.4\textwidth]{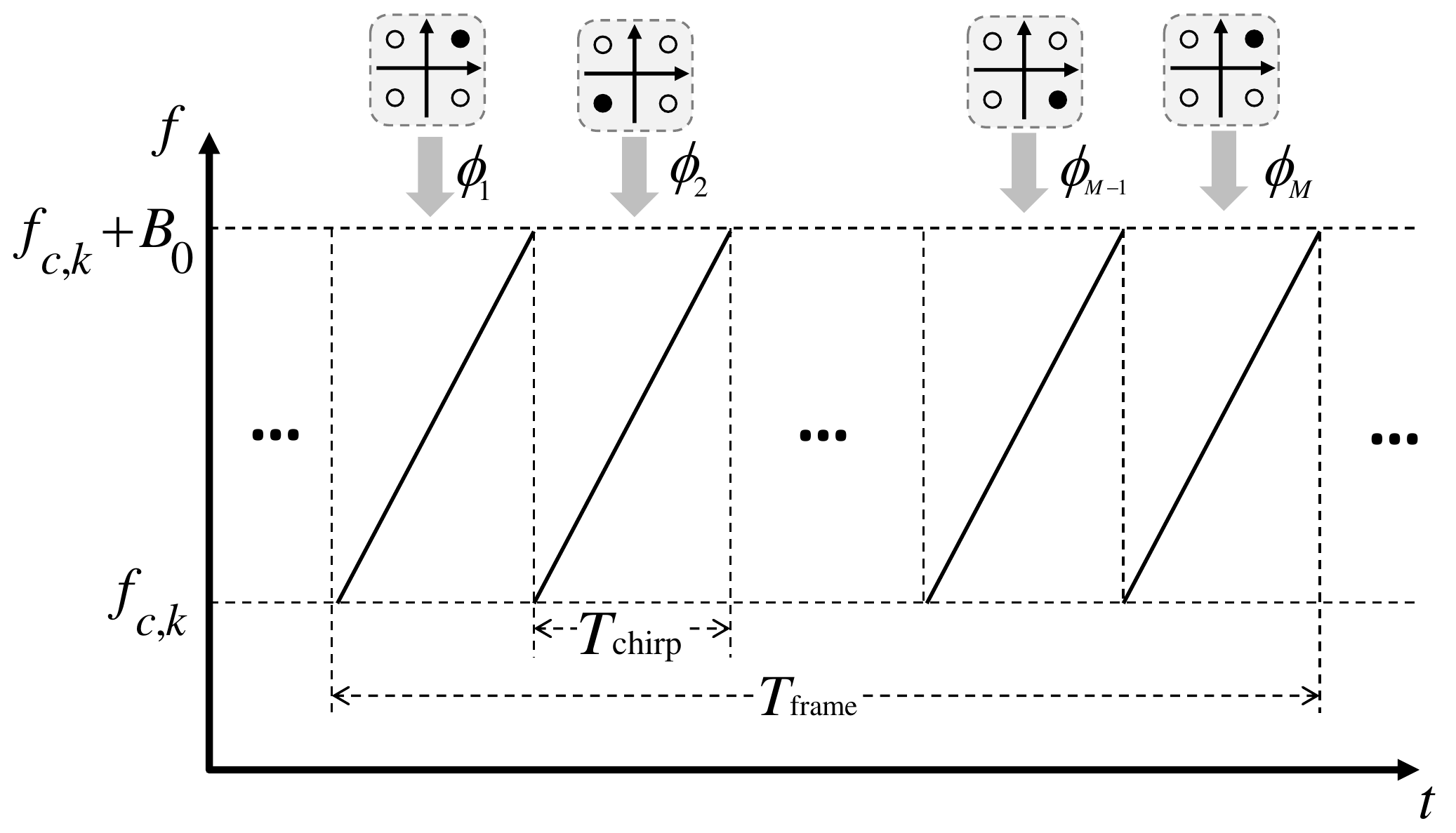}
	\caption{Basic frame structure of modulated FMCW signals.}
	\label{fig:frame}
\end{figure}
To avoid inter-device interference in the V-FEEL network, we assume that each edge device is assigned a dedicated sub-channel of equal bandwidth for ISAC. 
The bandwidth for each device is denoted as $B$, and its transmit power is denoted as $P_{T,k}$.
We adopt the modulated FMCW signals to enable the function of ISAC \cite{Oliveira2021TMTT}, as depicted in Fig. \ref{fig:frame}.
The frame of the ISAC signal comprises $M$ chirps each of time duration $T_{\text{chirp}}$, and the resulting frame duration is $T_{\text{frame}}=M T_{\text{chirp}}$.
Each of these chirps is a linear frequency up-ramp with a scope $\mu_{k} = B/T_{\text{chirp}}$ and modulated by symbols mapped onto a quadrature phase shift keying (QPSK) constellation. 
In other words, the information bits are modulated by altering the phase of transmit chirps. 
Accordingly, the transmit frame from edge device $k$ is given by
\begin{align*}
	x_{k}(t)&=\sum_{m=0}^{M-1}\text{rect}\left(\frac{t-mT_{\text{chirp}}}{T_{\text{chirp}}}\right) \\
	&\cdot \cos\left(2 \pi f_{c,k} \left(t-mT_{\text{chirp}}\right) + \pi \mu \left(t-mT_{\text{chirp}}\right)^{2} + \phi_{m}\right),
\end{align*}
where $f_{c,k} \gg B$ is the carrier frequency for device $k$, $\phi_{m} \in 
\lbrace \pi/4, 3\pi/4, 5\pi/4, 7\pi/4 \rbrace$ is the modulated phase of transmit chirp $m$ after QPSK modulation, and $\text{rect}(t)$ is the rectangular function that takes the value $1$ for $t\in[0,1)$ and $0$ otherwise. Moreover, the transmit signal is unmodulated when $\phi_{m} = 0$.

\color{black}
The received echo signal at edge device $k$ is given by
\begin{equation}\label{eq:sensing-signal}
	y_{k}^{\sf S}(t) = h_{k}^{\sf S}(t) \ast x_{k}(t) + e_{k}(t),\;\forall k \in \mathcal{K},
\end{equation}
where $\mathcal{K} \triangleq \{1,\ldots,K\}$ is the set of edge devices, $\ast$ denotes the convolution operation, and $e_{k}(t)$ is the additive noise term due to the ground clutter and the receiver noise. The ground clutter is caused by reflections from stationary objects in the environment. $h_{k}^{\sf S}(t)$ is the echo channel from the target, and its model depends on the specific sensing application, e.g., primitive based channel model for moving human \cite{Li2021SPAWC-sensing-model} or extended target model for points-like targets \cite{Leshem2007JSTSP-target}. 

The received uplink (UL) signal from device $k \in \overline{\mathcal{K}}\triangleq \{1,\ldots,K-1\}$ at device $K$  is given by
\begin{equation}\label{eq:comm-signal-K}
	y_{K}^{\sf C}(t) = \sum_{k \in \overline{\mathcal{K}}} h_{Kk}^{\sf C}(t) \ast x_{k}(t) + n_{K}(t),
\end{equation} 
and the received downlink (DL) signal from device $K$ at device $k \in \overline{\mathcal{K}}$ is given by
\begin{equation}\label{eq:comm-signal-k}
	y_{k}^{\sf C}(t) = h_{kK}^{\sf C}(t) \ast x_{K}(t) + n_{k}(t),\;\forall k \in \overline{\mathcal{K}},
\end{equation}
where $h_{kj}^{\sf C}(t)$ is the wireless communication channel from device $j$ to $k$. 
In this work, channel $h_{kj}^{\sf C}$ is modeled as the quasi-deterministic channel of IEEE 802.11ay \cite{Maltsev2016channel-ay}, i.e., 
\begin{align}\label{eq:comm-channel}
	\nonumber h_{kj}^{\sf C}(t) = & \sum_{n=1}^{N_{kj}} \sqrt{H_{n}}\frac{\lambda}{4\pi \left(D_{kj}+\tau_{n}^{\sf clustter}c\right)}\\
	&\times \sum_{m=1}^{M_{kj}}a_{n,m}\exp(j\psi_{n,m})\delta(t-\tau_{n,m}^{\sf ray}),
\end{align}
where $H_{n}$ is the reflection loss of the $n$-th cluster, $\lambda$ is the wave length, $D_{kj}$ is the distance between the $k$-th device and the $j$-th device, $c$ is the speed of light, $\tau_{n}^{\sf clustter}$ is the time delay of the $n$-th cluster, $a_{n,m}$, $\psi_{n,m}$, and $\tau_{n,m}^{\sf ray}$ are the amplitude, the initial phase, and the time delay of the $m$-th ray in the $n$-th cluster, respectively. $n_{k}(t)$ denotes the additive white Gaussian noise. 
\color{black}

\subsection{Signal Preprocessing at Receiver}\label{sec:sensing-processing}


\subsubsection{Sensing processing} Each edge device $k \in \mathcal{K}$ processes the received echo signal $y_{k}^{\sf S}(t)$ in \eqref{eq:sensing-signal} as follows. After down-conversion and bandpass filtering, the received echo signal is then fed into the A/D converter with sampling rate $F_{s}$.
Afterwards, the obtained complex vector is multiplied with the complex conjugate of the QPSK symbols used to modulate the corresponding transmit chirps. Then, the resulting vector in a frame is reconstructed as a sensing data matrix $\mathbf{Y}_{k} \in \mathbb{C}^{N_{c} \times M}$, where $N_{c} = T_{\text{chirp}}F_{s}$. In this matrix, each column vector represents a set of complex-valued baseband samples from a single chirp sampled at the rate $F_{s}$ along the fast-time axis, which can be used to extract range information; each row vector contains complex-valued baseband samples in $M$ different chirps from the same range bin, which can be used to extract Doppler information along the slow-time axis \cite{Liu2021arXiv-ISAC}. 

\subsubsection{Communication processing} \textcolor{black}{The communication channel $h_{kj}^{\sf C}$ in \eqref{eq:comm-channel} is assumed to be estimated based on a preamble, e.g., unmodulated chirps \cite{Oliveira2021TMTT}. To extract the necessary information for V-FEEL training and inference from the received ISAC signal $y_{k}^{\sf C}(t)$ in \eqref{eq:comm-signal-K} and \eqref{eq:comm-signal-k}, a matched-filter-like communication receiver similar to \cite{Oliveira2021TMTT} is adopted, by which the embedded QPSK symbols encoding the said information can be extracted and decoded.}
The resulting data rate is $\mathcal{R}^{C} = \frac{2}{T_{\text{chirp}}}$ as one chirp conveys 2 bits information due to the adopted QPSK modulation.
Note that during the training and inference process of the V-FEEL, the size of the transmitted vectors from each device is in the order of hundreds, small enough to yield the communication latency of only a few seconds. This will be validated by the simulation results in Section \ref{sec:results}.

\begin{remark}[Benefits of ISAC]
	Modulated FMCW signals as ISAC signals bring lower communication rate but better sensing performance compared to dedicated communication signals such as in cellular and WiFi systems.
	This fits right into our V-FEEL based DWS, where sensing is the priority and only relatively low communication rates are required, as each edge device only needs to transmit a vector with small size. 
	\color{black}Moreover, enabled by ISAC, all edge devices use the same transceiver for both communication and sensing, thus eliminating the need for deploying an additional central server that can only communicate as in a normal FEEL network.
	In general, our scheme can be self-organizing and scalable in practice.\color{black}
\end{remark}

\section{Vertical FEEL Scheme with DWS}\label{sec:VFL}

The section presents the details of the V-FEEL scheme. First, we describe the architecture of the V-FEEL framework and the problem formulation. Then, we elaborate the V-FEEL pipeline, including data generation, training stage, and inference stage.

\subsection{Data Generation}

Objects/human motion recognition relies on characterizing the micro-Doppler pattern embedded in the sensing data matrices $\mathbf{Y}_{k}$'s as introduced in Section \ref{sec:sensing-processing}. 
The latent discriminative features can be extracted from the micro-Doppler signatures in the spectrogram, which are then used for objects/human motion recognition \cite{Gurbuz2019SPM_SensingDL}. Therefore, we generate spectrograms as inputs of the V-FEEL as elaborated in the following.
We assume that the unit time requiring for sensing a motion is $T_{\text{spec}}$, and the resulting number of frames for sensing a motion is $N_{f}=T_{\text{spec}}/T_{\text{frame}}$. 
Then, the sensing data matrices in these $N_{f}$ frames at device $k$ can be concatenated as $\widetilde{\mathbf{Y}}_{k}=\left[\mathbf{Y}_{k}[1],\mathbf{Y}_{k}[2],....,\mathbf{Y}_{k}[N_{f}]\right] \in \mathbb{C}^{N_{c} \times N_{f}M}$. 
Following \cite{Li2021SPAWC-sensing-model}, $\widetilde{\mathbf{Y}}_{k}$ is then passed through a filter based on singular value decomposition (SVD) to remove the ground clutter, yielding a denoised data matrix $\overline{\mathbf{Y}}_{k}$. 
Then, the data matrix $\overline{\mathbf{Y}}_{k}$ is converted to a vector $\overline{\mathbf{y}}_{k} \in \mathbb{C}^{N_{f}M}$ by summing up the data along the fast-time dimension, followed by short-time Fourier transform (STFT) to generate a spectrogram. 

\color{black}
It is remarked that the micro-Doppler signature depends on the cosine of the angle between the trajectory of the target and the sensor line-of-sight (aspect angle).
Hence, some discriminative features may not be observable from certain aspect angles at some sensors, but can be easily resolved from other aspect angles at other sensors \cite{Fioranelli2017mD-feature}.
For example, in human motion recognition, sensors located on the side of the body cannot obtain the micro-Doppler signatures of the frontal parts of the body.
In this regard, the latent discriminative feature spaces embedded in the spectrograms at different sensors are generally distinct (though possibly with  partial overlapping). This well fits the scenarios aimed by the V-FEEL.
\color{black}

\begin{figure}[!t]
	\centering
	\includegraphics[width=0.45\textwidth]{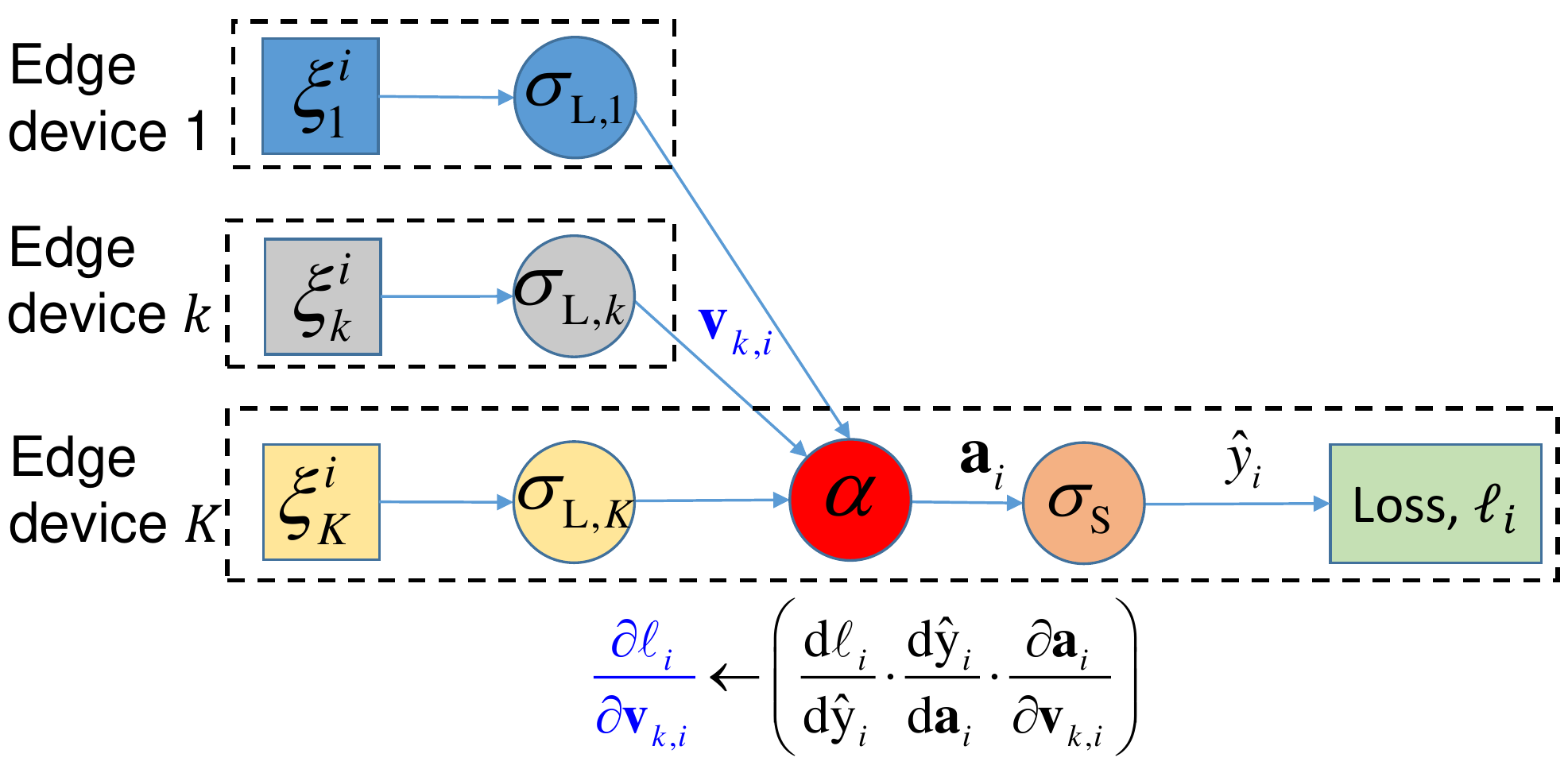}
	\caption{Architecture and model training process of vertical FEEL. In the forward propagation, $\mathbf{v}_{k,i}$ will be sent to  device $K$ from device $k$ via ISAC transmission; in the back-propagation, $\frac{\partial \ell_{i}}{\partial \mathbf{v}_{k,i}}$ will be sent to device $k$ from device $K$ via ISAC transmission.}
	\label{fig:VFL}
\end{figure}

\subsection{Architecture and Problem Formulation}
The architecture of our adopted V-FEEL framework is shown in Fig. \ref{fig:VFL}. Let $\mathcal{D}_{i} = \left\lbrace (\xi_{1,i},\xi_{2,i},...,\xi_{K,i}), y_{i} \right\rbrace$ represent the $i$-th training sample, where $\xi_{k,i}$ is the sample generated at the $k$-th edge device, and $y_{i}$ is the label in  the $i$-th training sample. 
Without loss of generality, we assume that the selected coordinator, i.e., device $K$, holds the labels. 
\textcolor{black}{We define two kinds of sub-models named by L-model and S-model, which can be neural networks or other machine learning models.} An individual local L-model is owned by each device to transform its input into an intermediate vector, and a downstream common S-model is owned only by the coordinator (device $K$) to calculate the final output.
The L-model owned by device $k$ is represented as a parametric function $\sigma_{{\sf L},k}(\cdot;\mathbf{w}_{{\sf L},k})$ with trainable parameter vector $\mathbf{w}_{{\sf L},k}$, while the S-model is represented as a parametric function $\sigma_{\sf S}(\cdot;\mathbf{w}_{{\sf S}})$ with trainable parameter vector $\mathbf{w}_{{\sf S}}$. 
The function $\alpha(\cdot)$ at device $K$ is used for aggregating the intermediate vectors from all the L-models. Typical aggregation functions can be element-wise average and concatenation \cite{Ceballos2020VFL-splitNN}. In our scheme, the objective is to learn all the L-models and the S-model to fit all data across the edge devices. In particular, we aim to solve:
\begin{equation}\label{eq:problem}
	\underset{\lbrace \mathbf{w}_{{\sf L},k} \rbrace, \mathbf{w}_{{\sf S}}}{\min} L = \frac{1}{N} \sum_{i=1}^{N} \ell(\lbrace \mathbf{w}_{{\sf L},k} \rbrace, \mathbf{w}_{{\sf S}};\mathcal{D}_{i}),
\end{equation}
where $\ell(\lbrace \mathbf{w}_{{\sf L},k} \rbrace, \mathbf{w}_{{\sf S}};\mathcal{D}_{i})$ is the loss function for the $i$-th sample, and $N$ is the number of training samples. 

\subsection{Training Stage}

The distributed stochastic gradient decent (SGD) algorithm is designed to solve problem (\ref{eq:problem}) in the V-FEEL. The training process involves multiple iteration, each of which includes both the forward propagation for loss function evaluation and the backward propagation for calculating the gradient. Unlike in centralized learning or H-FEEL where the entire computation graph is computed at each edge device, in V-FEEL each edge device keeps only a portion of the computation graph, and the intermediate results must be exchanged between each edge device $k$ and the coordinator (device $K$) in the course of forward propagation and back-propagation.

For any iteration, say the $r$-th iteration, the L-model owned by each edge device maps their data samples $\xi_{k,i}^{(r)}$ to an intermediate vector $\mathbf{v}_{k,i}^{(r)} = \sigma_{{\sf L},k}(\xi_{k,i}^{(r)};\mathbf{w}_{{\sf L},k}^{(r)}) \in \mathbb{R}^{d}$.
Then each edge device $k \in \lbrace 1,2,...,K-1 \rbrace$ transmits its output intermediate vector $\mathbf{v}_{k,i}^{(r)}$ to device $K$ via ISAC transmission. 
After receiving intermediate vectors from the other $K-1$ devices, the coordinator (device $K$) aggregates with its own output $\mathbf{v}_{K,i}^{(r)}$ to obtain vector $\mathbf{a}_{i}^{(r)} = \alpha(\mathbf{v}_{1,i}^{(r)},\mathbf{v}_{2,i}^{(r)},...,\mathbf{v}_{K,i}^{(r)})$.
The aggregated result is then passed through the S-model, and the predicted output for $i$-th training sample is $\hat{y}_{i}^{(r)} = \sigma_{\sf S}(\mathbf{a}_{i}^{(r)};\mathbf{w}_{{\sf S}}^{(r)})$. 
Finally, the loss for $i$-th sample can be calculated as $\ell_{i}^{(r)} = \epsilon(\hat{y}_{i}^{(r)},y_{i})$ where $\epsilon(\cdot,\cdot)$ is the error function, e.g., the cross entropy loss, between the actual output and the predicted output. 
To sum up, the loss for $i$-th sample is given by
\[
\ell_{i}^{(r)} = \epsilon\left(\sigma_{\sf S}\left(\alpha\left(\lbrace \sigma_{{\sf L},k}(\xi_{k,i}^{(r)};\mathbf{w}_{{\sf L},k}^{(r)}) \rbrace\right);\mathbf{w}_{{\sf S}}^{(r)}\right),y_{i}^{(r)}\right).
\]
\color{black}
Once the sample loss is obtained, device $K$ performs back-propagation over S-model $\sigma_{\sf S}$ and obtains the stochastic gradient of the loss function $L$ with respect to $\mathbf{w}_{{\sf S}}^{(r)}$, denoted as $\tilde{\nabla} L(\mathbf{w}_{{\sf S}}^{(r)})$. Besides, the partial derivative of the loss with respect to each intermediate vector $\mathbf{v}_{k,i}^{(r)}$ can also be calculated by using the chain rule:
\begin{equation*}
	\frac{\partial \ell_{i}^{(r)}}{\partial \mathbf{v}_{k,i}^{(r)}} = \frac{\texttt{d} \ell_{i}^{(r)}}{\texttt{d} \hat{y}_{i}^{(r)}} \cdot \frac{\texttt{d} \hat{y}_{i}^{(r)}}{\texttt{d} \mathbf{a}_{i}^{(r)}} \cdot \frac{\partial \mathbf{a}_{i}^{(r)}}{\partial \mathbf{v}_{k,i}^{(r)}}.
\end{equation*}
Device $K$ sends each $\partial \ell_{i}^{(r)}/\partial \mathbf{v}_{k,i}^{(r)}$ to the other devices via ISAC transmission, and each device can perform its respective back-propagation to compute the stochastic gradient of $L$ with respect to $\mathbf{w}_{{\sf L},k}^{(r)}$, denoted as $\tilde{\nabla} L(\mathbf{w}_{{\sf L},k}^{(r)})$.
\color{black}

After back-propagation, the stochastic gradients, i.e.,  $\tilde{\nabla} L(\mathbf{w}_{{\sf S}}^{(r)})$ and $\tilde{\nabla} L(\mathbf{w}_{{\sf L},k}^{(r)})$, can be used for updating the S-model and L-models, respectively:
\begin{align*}
	\mathbf{w}_{{\sf S}}^{(r+1)} &= \mathbf{w}_{{\sf S}}^{(r)} - \eta_{\sf S}^{r} \tilde{\nabla} L(\mathbf{w}_{{\sf S}}^{(r)}),\\
	\mathbf{w}_{{\sf L},k}^{(r+1)} &= \mathbf{w}_{{\sf L},k}^{(r)} - \eta_{{\sf L},k}^{r} \tilde{\nabla} L(\mathbf{w}_{{\sf L},k}^{(r)}),\;\forall k \in \lbrace 1,2,...,K \rbrace,
\end{align*}
where $\eta_{\sf S}^{r}$ and $\eta_{{\sf L},k}^{r}$ are the corresponding learning rates.

Upon the training stage is completed, the well-trained L-models and S-model are deployed for inference, and the process is similar to forward propagation in the training stage.

\section{Performance Evaluations}\label{sec:results}

This section evaluates the performance of the proposed V-FEEL based DWS via a human motion recognition example.

\subsection{Simulation Setup}

\begin{figure}[!t]
	\centering
	\includegraphics[width=0.32\textwidth]{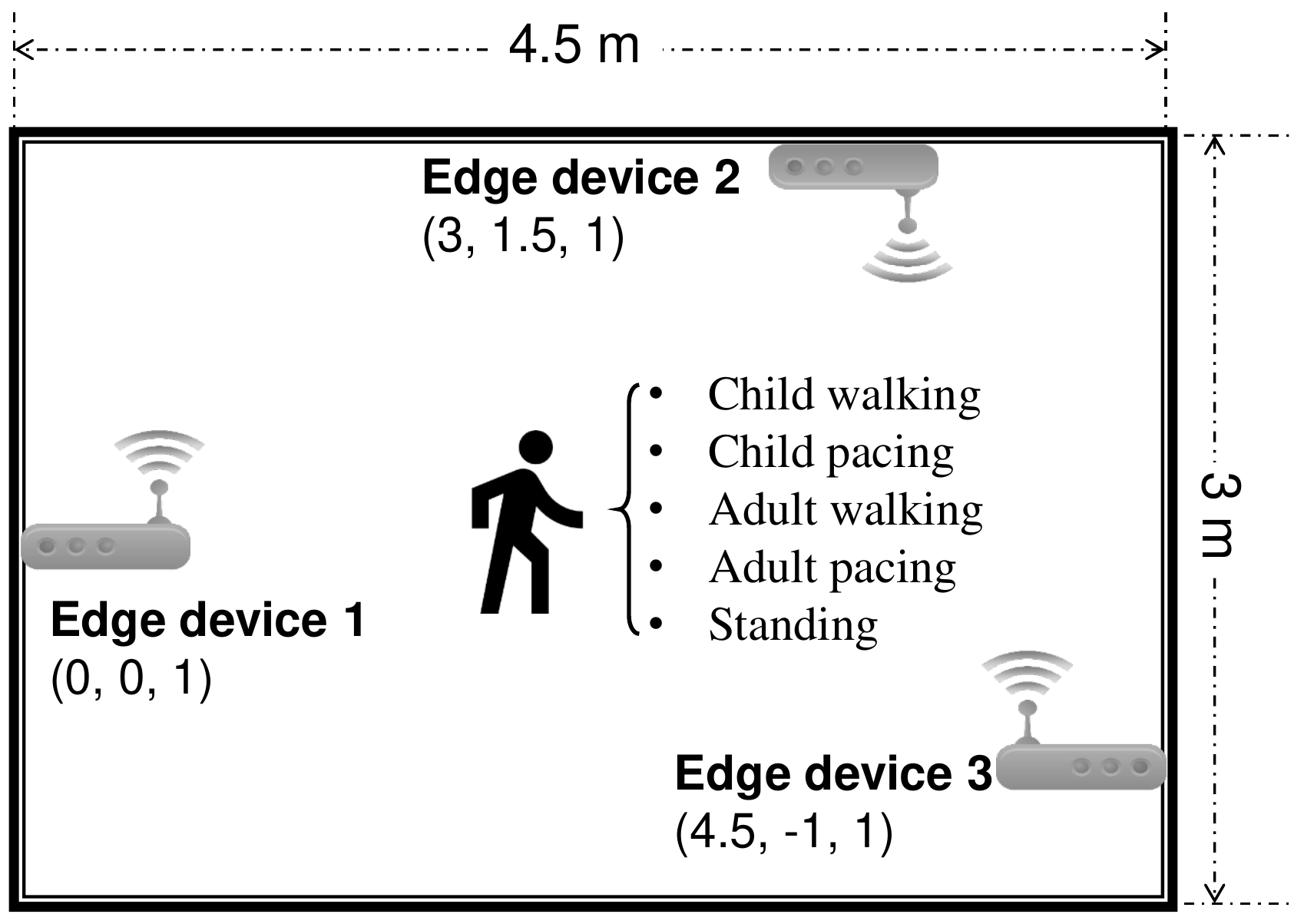}
	\caption{Illustration of the simulation scenario.}
	\label{fig:sim-scenario}
\end{figure}

\begin{figure}[!t]
	\centering
	\includegraphics[width=0.48\textwidth]{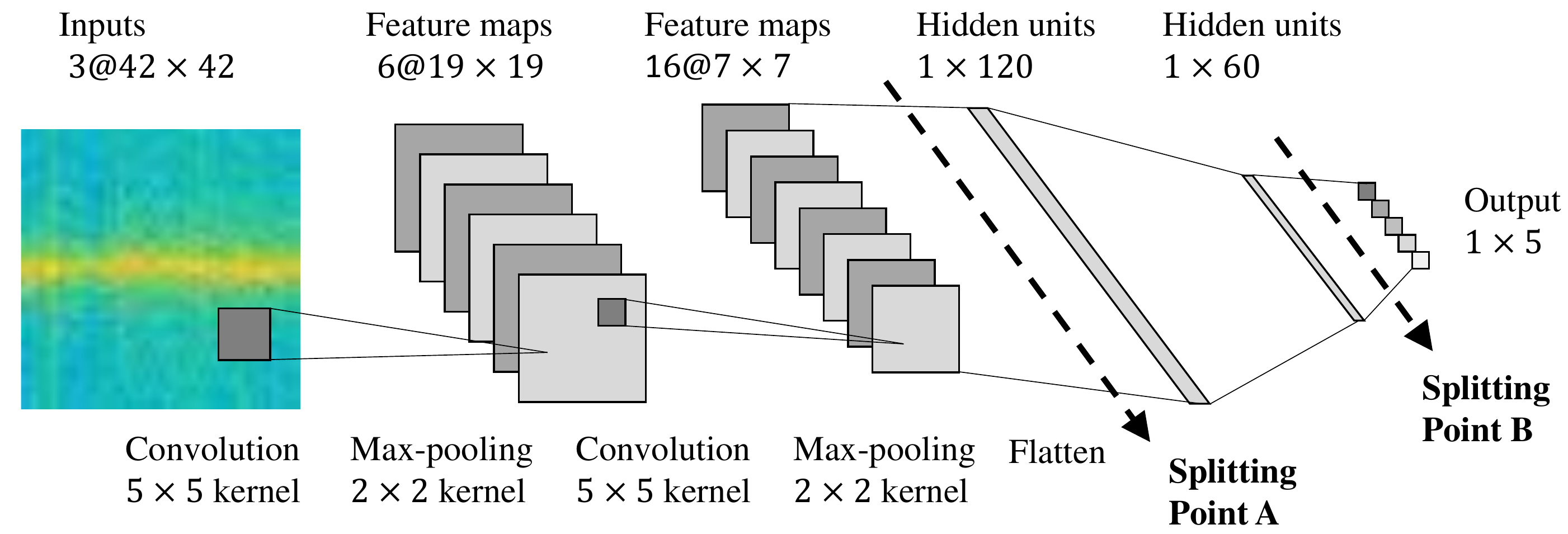}
	\caption{Structure of the complete CNN model.}
	\label{fig:CNN}
\end{figure}


\begin{table}[!t]
	\renewcommand{\arraystretch}{1.2}
	\caption{Simulation Parameters for Human Motion Recognition}
	\label{tab:sensing-parameters}
	\centering
	\begin{tabular}{l l l}
		\hline
		\bfseries Parameter & \bfseries Notation & \bfseries Value\\
		\hline\hline
		Bandwidth & $B$ & $10$MHz \\
		Carrier frequency & $f_{c,1}$, $f_{c,2}$, $f_{c,3}$ & $60$, $60.01$, $60.02$ GHz\\
		Chirp duration & $T_{\text{chirp}}$ & $10$$\mu s$ \\
		Chirp number per frame & $M$ & 25 \\
		Unit time for sensing & $T_{\text{spec}}$ & $0.5$$s$ \\
		Sampling rate & $F_{s}$ & $10$MHz \\
		Transmit power & $P_{T,k}$ & $1$W\\
		\hline
	\end{tabular}
\end{table}

\subsubsection{Human motion recognition} We apply the human motion recognition platform in \cite{Li2021SPAWC-sensing-model} to simulate various human motions and generate human motion datasets.
In our simulation, the sensing task is to identify five different human motions, i.e., \textit{child walking}, \textit{child pacing}, \textit{adult walking}, \textit{adult pacing}, and \textit{standing}. The heights of children and adults are uniformly distributed in interval $[0.9\text{m}, 1.2\text{m}]$ and $[1.6\text{m}, 1.9\text{m}]$, respectively. The speed of standing, walking, and pacing are $0$ m/s, $0.5H$ m/s, and $0.25H$ m/s, respectively, where $H$ is the height value.
As shown in Fig. \ref{fig:sim-scenario}, the room size is $4.5$m (length) $\times$ $3$m (width) $\times$ $3$m (height). $K=3$ edge devices are distributed around the room. We set the coordinate of the first edge device as $(0\text{m}, 0\text{m}, 1\text{m})$, and the coordinates of other two devices as $(3\text{m}, 1.5\text{m}, 1\text{m})$ and $(4.5\text{m}, -1\text{m}, 1\text{m})$, respectively. Moreover, we set the initial moving position of each human motion in a central rectangular area with size $3$m $\times$ $2$m to prevent the objects from moving outside the room. The heading of each object is set to be uniformly distributed in $[–180^{\circ}, 180^{\circ}]$. The simulation parameters are listed in Table \ref{tab:sensing-parameters}.

\subsubsection{Vertical FEEL} Since the sample sizes of each device are the same, we consider that all the devices hold the same L-model. \color{black}
In this experiment, to generate L-model and S-model, we split a $6$-layer CNN on a per-layer basis as shown in Fig. \ref{fig:CNN}. 
The front half containing the input layer is selected as L-model, and the end half containing the classification layer is selected as S-model. Two splitting schemes, i.e., splitting at Splitting Point A or Splitting Point B, are evaluated. Splitting at Splitting Point A generates small L-model and large S-model; splitting at Splitting Point B generates large L-model and small S-model. In addition, the size of the intermediate vectors generated by splitting at Splitting Point A and Splitting Point B are 784 and 60, respectively. 
\color{black}
We also evaluate two aggregation schemes for aggregating the intermediate vectors, i.e., element-wise average (EWA) or concatenation (CAT). Different combinations of splitting schemes and aggregation schemes lead to four different V-FEEL schemes, which are named as VFL-A-EWA, VFL-A-CAT, VFL-B-EWA, and VFL-B-CAT, respectively. The learning rates of all sub-models are set to be $0.01$, and the mini-batch size is set to be 32.



\subsection{Performance of V-FEEL based DWS}

%



\begin{figure}[!t]
	\begin{minipage}{0.23\textwidth}
		\includegraphics[width=\textwidth]{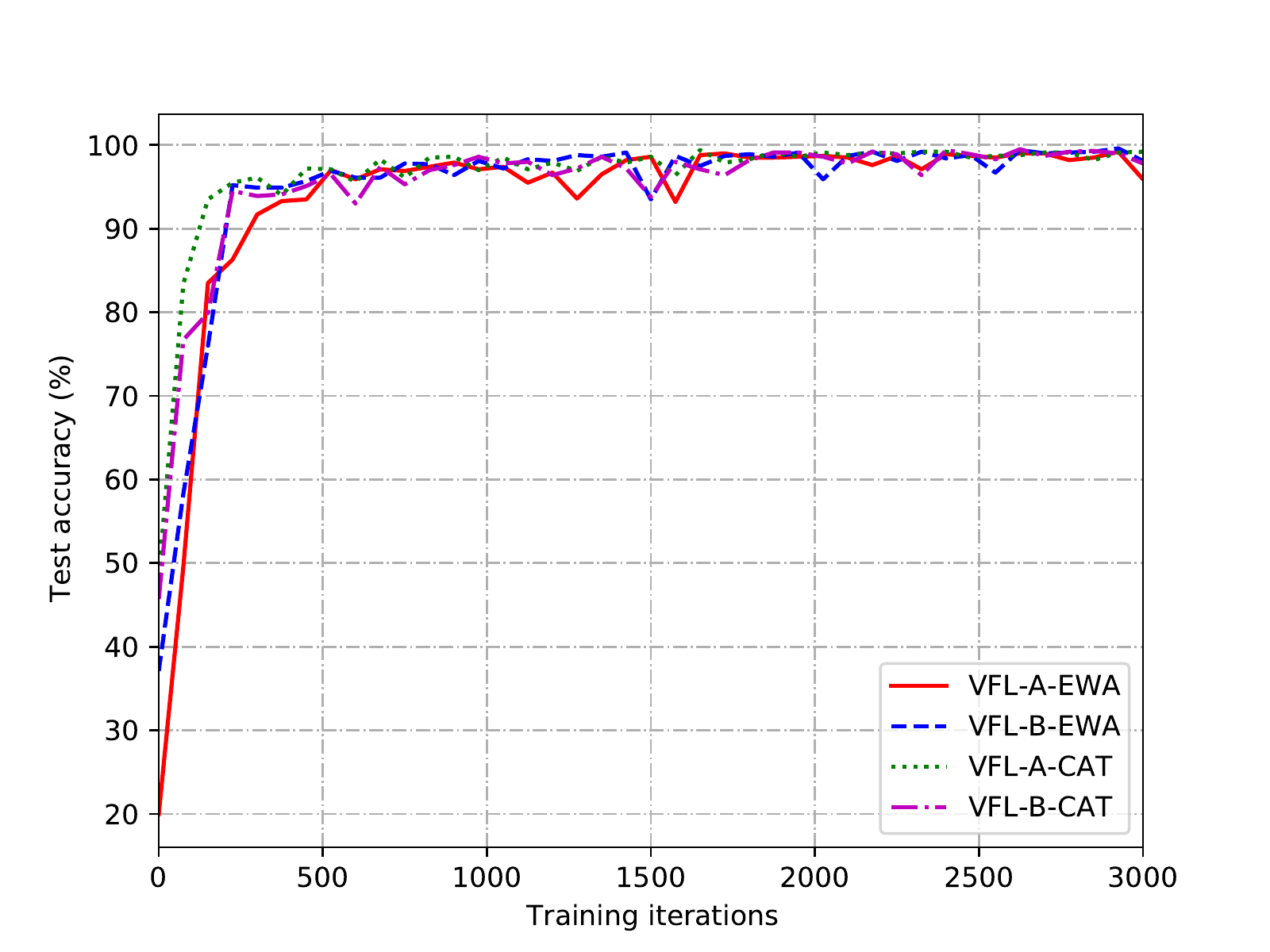}
		\vspace*{-5mm}
		\caption{Test accuracy versus iteration of different V-FEEL schemes.}
		\label{fig_acc}
	\end{minipage}\hfill
	\begin{minipage}{0.23\textwidth}
		\includegraphics[width=\textwidth]{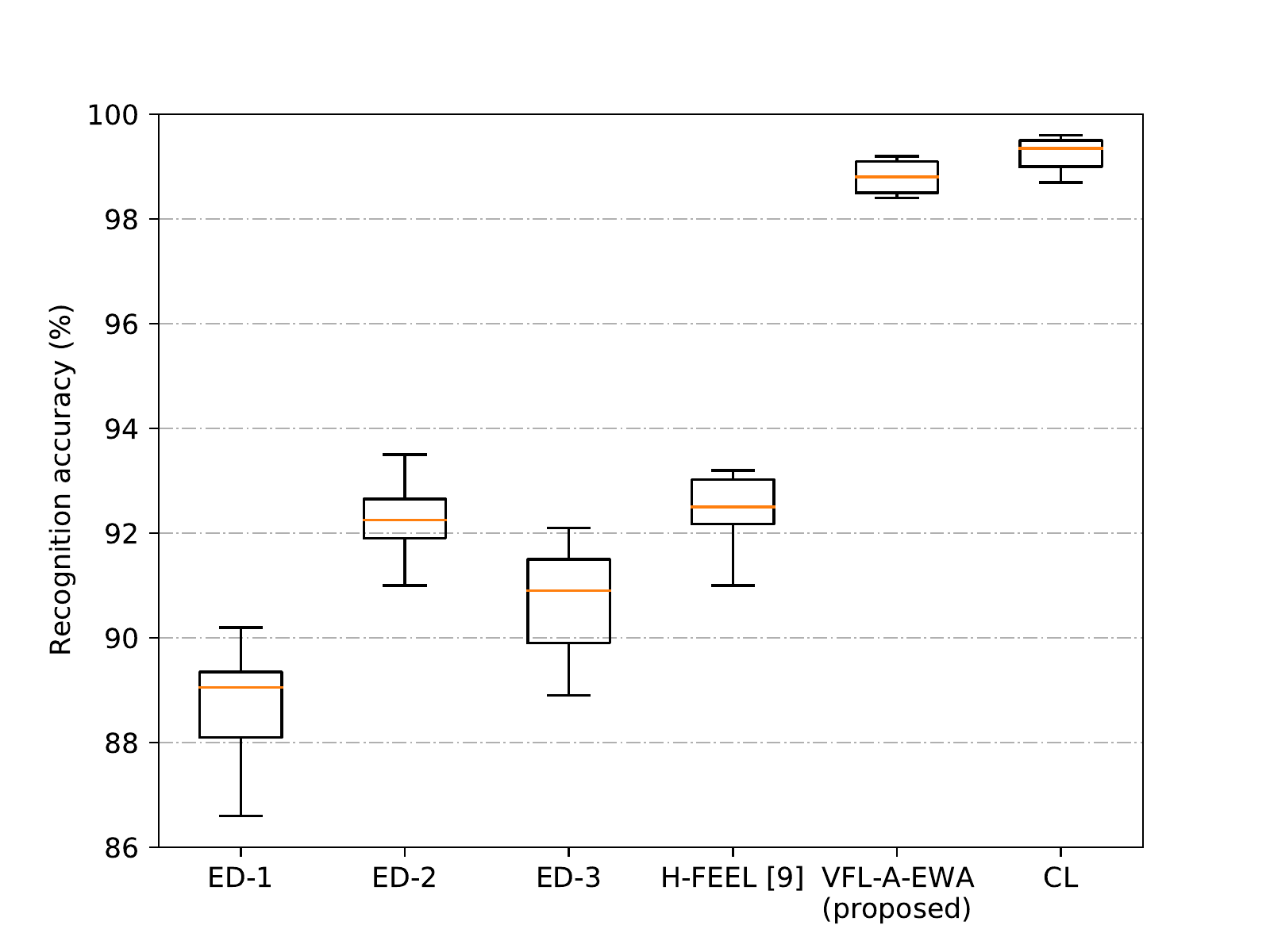}
		\vspace*{-5mm}
		\caption{Recognition accuracy with different learning schemes.}
		\label{fig:box}
	\end{minipage}
\end{figure}

First, we evaluate the four V-FEEL schemes in terms of test accuracy versus training iteration, as shown in Fig. \ref{fig_acc}. \textcolor{black}{We observe that the four schemes achieve similar performances, suggesting the performance of the V-FEEL is to some extent robust against the different splitting and aggregation schemes, although the optimal design remains an open problem.}
Thus, without loss of generality, we use the scheme of VFL-A-EWA to evaluate the performance of the V-FEEL in the following.

Next, we compare the recognition accuracy of models trained by six schemes: ED-$k$ ($k$=1, 2, 3), H-FEEL \cite{Yang2022FL-multi-radar}, centralized learning (CL), and VFL-A-EWA, where ED-$k$ represents on-device training at edge device $k$.
In CL, we concatenate the spectrograms from all the edge devices for the same motion to form a new sample. Fig. \ref{fig:box} depicts the box plot of recognition accuracy with different schemes, which shows the minimum, the maximum, the sample median, and the first and third quartiles of recognition accuracies collected in 12 experiments for each scheme. As seen, CL, H-FEEL and VFL-A-EWA achieve higher recognition accuracy than ED-$k$. It is because that the models from CL, H-FEEL and VFL-A-EWA are trained cooperatively by three edge devices, thus exploiting more data in the training process. Moreover, CL achieves the highest recognition accuracy, but needs to share the raw data among the edge devices, and thus leading to severe data privacy issue. We also observe that VFL-A-EWA achieves similar performance with CL, but performs much better than H-FEEL due to its better exploitation of the multi-view nature of sensing data by taking the sensing data from different views of a common target as a part of entire feature space instead of independent samples as did in H-FEEL. 

\subsection{Communication and Computation Overhaed}

\begin{table}[!t]
	\renewcommand{\arraystretch}{1.3}
	\caption{Communication and Computation Overhead of V-FEEL and H-FEEL}
	\label{tab:c2-analysis}
	\centering
	\begin{tabular}{lllll}
		\hline
		\multicolumn{2}{l}{}                         &\bfseries VFL-A-EWA   &\bfseries VFL-B-EWA    &\bfseries H-FEEL                     \\ \hline\hline
		\multirow{2}{*}{Model size}     & L-model    & 2,872       & 104,322      & \multirow{2}{*}{104637} \\ \cline{2-4}
		& S-model    & 101,765     & 305          &                         \\ \hline
		\multicolumn{2}{l}{GFLOPs/Iteration}             & \textbf{0.1782}       & 0.1847        & 0.1847                   \\ \hline
		\multicolumn{2}{l}{Comm. time (s)/Iteration} & 4.27 & \textbf{0.33  } & 33.48             \\ \hline
	\end{tabular}
\end{table}

Table \ref{tab:c2-analysis} demonstrates the communication overhead and computation flops in VFL-A-EWA, VFL-B-EWA, and H-FEEL in an iteration. The sizes of L-model and S-model depend on the splitting point. 
The computation flops in the V-FEEL can be calculated as $C_{\sf L}K + C_{\sf S}$, where $C_{\sf L}$ and $C_{\sf S}$ are the computation flops for L-model and S-model, respectively. We see that VFL-A-EWA consumes less flops than VFL-B-EWA, since it has a smaller L-model and a larger S-model than VFL-B-EWA. On the other hand, the computation flops in H-FEEL is calculated as $K(C_{\sf L} + C_{\sf S})$. In VFL-B-EWA, the computation flops for L-model is much higher than that for S-model, so it holds that $C_{\sf L}K + C_{\sf S} \approx K(C_{\sf L} + C_{\sf S})$.
Thus, we can see that VFL-B-EWA and H-FEEL have similar flop consumption.
As for communication overhead, the communication time in VFL-A-EWA is much shorter than that in VFL-B-EWA, since the size of intermediate vectors to be transmitted in VFL-A-EWA is much smaller. In H-FEEL, each edge device has to transmit all the whole model parameters, so the communication time is much longer.
\color{black}
In general, the communication and computation overheads of different V-FEEL schemes depend on the architectures of L-models and S-model. 
Thus, our V-FEEL scheme can adapt to the computing powers and communication capacities of the edge devices in the network through flexible model splitting.
\color{black}
Moreover, the superiority in communication overhead of V-FEEL over H-FEEL is pronounced.

\section{Conclusion}

In this letter, we proposed a privacy-preserving vertical FEEL scheme for collaborative objects/human motion recognition by exploiting distributed ISAC.
Extensive experimental results are provided to confirm the performance gained through our proposed scheme. 
In light of these results, we believe that our proposed scheme can also contribute to various wireless sensing related applications, such as Internet of Things (IoT) and healthcare delivery.
Future work will entail optimizing vertical FEEL structure to accommodate different learning tasks and edge devices with heterogeneous communication and computation capacities, and taking into account the multimodal data generated by different kinds of sensors, etc.

\bibliographystyle{IEEEtran}
\bibliography{VFL}
\end{document}